\documentclass[twocolumn,nopacs,preprintnumbers,amsmath,amssymb,reprint,prl]{revtex4-1}
\usepackage{epsfig}
\usepackage{dcolumn}
\usepackage{bm}

\newcommand{\remove}[1]{}

\def\be{\begin{equation}}
\def\ee{\end{equation}}

\newcommand{\beq}{\begin{equation}}
\newcommand{\eeq}{\end{equation}}
\newcommand{\beqa}{\begin{eqnarray}}
\newcommand{\eeqa}{\end{eqnarray}}

\newcommand{\ii}{{\rm i}}

\newcommand{\vv}{{\bf v}}

\newcommand{\vx}{{\bf x}}
\newcommand{\vk}{{\bf k}}
\newcommand{\vp}{{\bf p}}
\newcommand{\vq}{{\bf q}}

\newcommand{\vF}{{\bf F}}

\newcommand{\tdelta}{{\tilde{\delta}}}

\newcommand{\tlambda}{{\tilde{\lambda}}}

\newcommand{\tC}{{\tilde{C}}}
\newcommand{\tR}{{\tilde{R}}}

\newcommand{\tvv}{{\tilde{\bf v}}}
\newcommand{\tvp}{{\tilde{\bf p}}}

\newcommand{\cD}{{\cal D}}

\newcommand{\bea}{\begin{array}}
\newcommand{\ea}{\end{array}}

\begin{document}
\title{Nonzero Density-Velocity Consistency Relations for Large Scale Structures}
\author{Luca Alberto Rizzo $^{1,2,3}$, David F. Mota $^{3}$, Patrick Valageas $^{1,2}$}
\affiliation{$^1$ Institut de Physique Th\'eorique,
CEA, IPhT, F-91191 Gif-sur-Yvette, C\'edex, France}
\affiliation{$^2$ Institut de Physique Th\'eorique,
CNRS, URA 2306, F-91191 Gif-sur-Yvette, C\'edex, France}
\affiliation{$^3$ Institute of Theoretical Astrophysics, University of Oslo, 0315 Oslo, Norway}


%
\begin{abstract}
We present exact kinematic consistency relations for cosmological structures that do not vanish 
at equal times and can thus be measured
in surveys. These rely on cross-correlations between the density and
velocity, or momentum, fields. Indeed, the uniform transport of small-scale structures
by long-wavelength modes, which cannot be detected at equal times by looking
at density correlations only, gives rise to a shift in the amplitude of the velocity field
that could be measured. 
These consistency relations only rely on the weak equivalence principle and Gaussian initial
conditions. They remain valid in the nonlinear regime and for biased galaxy fields.
They can be used to constrain nonstandard cosmological scenarios or the large-scale
galaxy bias.
%
\end{abstract}
%
%
\maketitle
\textit{Introduction:}
\label{introduction}
Cosmological structures can be described on large scales
by perturbative methods while smaller scales are described by phenomenological models
or studied with numerical simulations. This makes it difficult to obtain accurate predictions
on the full range of scales probed by galaxy or weak lensing surveys.
Moreover, if we consider galaxy density fields, theoretical predictions remain sensitive 
to the galaxy bias (galaxies do not exactly follow the matter density field),
which involves some phenomenological modeling of star formation.

This makes exact analytical results that go beyond low-order perturbation theory
and apply to biased tracers very rare. However, such exact results have 
recently been obtained 
\cite{Kehagias2013,Peloso2013,Creminelli2013,Kehagias2014c,Peloso2014,Creminelli2014a,Valageas2014a,Horn2014,Horn2015}
in the form of ``kinematic consistency relations''. 
They relate the $(\ell+n)$-density (or velocity divergence) correlations, 
with $\ell$ large-scale wave numbers
and $n$ small-scale wave numbers, to the $n$-point small-scale correlation.
These relations, obtained at the leading order over the large-scale wave numbers,
arise from the equivalence principle. It ensures
that small-scale structures respond to a large-scale perturbation (which at leading
order corresponds to a constant gravitational force over the extent of the small-sized
object) by a uniform displacement. Therefore, these relations express
a kinematic effect that vanishes for equal-time statistics, 
as a uniform displacement has no impact on the statistical properties of the density
field observed at a given time.

In practice, it is difficult to measure different-time correlations and it is useful 
to obtain relations that remain nonzero at equal times.
This is possible by going to the next order and taking into account tidal effects,
which at their leading order are given by the response of small-scale structures to a change 
of the background density. However, in order to derive expressions that apply to our Universe
one needs to introduce some additional approximations 
\cite{Valageas2014b,Kehagias2014,Nishimichi2014}.

In this Letter, we show that it is possible to derive exact kinematic consistency relations
that do not vanish at equal times by considering cross-correlations between the density
and velocity, or momentum, fields.
Indeed, the uniform displacement due to the long-wavelength mode also gives rise to a shift
in the amplitude of the velocity field that does not vanish at equal times and can thus be 
observed. These consistency relations have the same degree of validity as the
previously derived density (or velocity divergence) relations and only rely on the 
weak equivalence principle and Gaussian initial conditions.

\textit{Correlation and response functions:}
%
The consistency relations that apply to large-scale structures assume that the system
is fully defined by Gaussian initial conditions (the primordial fluctuations that are found at the
end of the inflationary epoch).
Thus, the dynamics is fully determined by the Gaussian linear matter growing mode 
$\delta_{L0}(\vx)$ (which we normalize today as usual) that directly
maps the initial conditions and can be observed on very large linear scales.
Then, any dependent quantities $\{\rho_1,...,\rho_n\}$, such as the dark matter or galaxy
densities at space-time positions $(\vx_i,\tau_i)$, are functionals of the field
$\delta_{L0}(\vx)$ and we can write the mixed correlation functions 
over $\delta_{L0}$ and $\{\rho_i\}$ as Gaussian averages, 
\beqa
C^{1,n}(\vx) & = & \langle \delta_{L0}(\vx)  \rho_1 \dots \rho_n \rangle  \nonumber \\ 
& = & \int {\cal{D}} \delta_{L0} \; 
e^{-\delta_{L0} \cdot C_{L0}^{-1} \cdot \delta_{L0}/2} \; 
\delta_{L0}(\vx) \rho_1 \dots  \rho_n , \;\;\;
\label{mixed_correlation}
\eeqa
where $C_{L0}(\vx_1,\vx_2) = \langle \delta_{L0}(\vx_1) \delta_{L0}(\vx_2) \rangle$
is the two-point correlation function of the Gaussian field $\delta_{L0}$.
Integrating by parts over $\delta_{L0}$ gives
\beq
C^{1,n}(\vx) = \int d\vx' \; C_{L0}(\vx,\vx') \, R^{1,n}(\vx') ,
\label{C1n-R1n}
\eeq
where we introduced the mean response function
\beq
R^{1,n}(\vx) = \langle \frac{\cD [ \rho_1 \dots  \rho_n ]}{\cD \delta_{L0}(\vx)} \rangle.
\label{R1n-def}
\eeq

Equation (\ref{C1n-R1n}) describes how the mixed correlation (\ref{mixed_correlation})
between the initial Gaussian field $\delta_{L0}$ and the dependent quantities $\{\rho_i\}$
is related to the response function of the latter to this Gaussian field.
Going to Fourier space, which we denote with a tilde, with the normalizations 
$\delta_{L0}(\vx)=\int d\vk \, e^{\ii\vk\cdot\vx} \tdelta_{L0}(\vk)$ and
$\langle \tdelta_{L0}(\vk_1) \tdelta_{L0}(\vk_2) \rangle = P_{L0}(k_1) \delta_D(\vk_1+\vk_2)$, 
equation (\ref{C1n-R1n}) gives
\beq
\tC^{1,n}(\vk) = P_{L0}(k) \, \tR^{1,n}(-\vk) ,
\label{C1n-R1n-Fourier}
\eeq
where we defined the Fourier-space correlation and response functions as
\beq
\tC^{1,n}(\vk) = \langle \tdelta_{L0}(\vk) \rho_1 \dots \rho_n \rangle , \;\;\;
\tR^{1,n}(\vk) = \langle \frac{\cD [ \rho_1 \dots  \rho_n ]}{\cD \tdelta_{L0}(\vk)} \rangle .
\nonumber
\eeq

\textit{Consistency relations for the density contrast:}
%
If we consider the quantities $\{\rho_i\}$ to be the nonlinear matter density contrasts 
$\tdelta(\vk_i,\tau_i)$ at wave number $\vk_i$ and conformal time $\tau_i$,
equation (\ref{C1n-R1n-Fourier}) writes as
\beqa
\langle \tdelta_{L0}(\vk') \tdelta(\vk_1,\tau_1) \dots \tdelta(\vk_n,\tau_n) \rangle & = &
P_{L0}(k') \nonumber \\
&& \hspace{-2.3cm} \times \langle \frac{\cD[ \tdelta(\vk_1,\tau_1) \dots \tdelta(\vk_n,\tau_n) ]}
{\cD \tdelta_{L0}(-\vk')} \rangle . \;\;
\label{consistency-delta-1}
\eeqa
On large scales the density field is within the linear regime, 
$\tdelta(\vk',\tau') \rightarrow D_+(\tau') \tdelta_{L0}(\vk')$; then for $k'\rightarrow 0$
\beqa
k'\rightarrow 0 : \;\;\; \langle \tdelta(\vk',\tau') \tdelta(\vk_1,\tau_1) \dots \tdelta(\vk_n,\tau_n) 
\rangle & = & \nonumber \\
&& \hspace{-5.5cm} D_+(\tau') P_{L0}(k') \langle \frac{\cD[ \tdelta(\vk_1,\tau_1) \dots 
\tdelta(\vk_n,\tau_n) ]}{\cD \tdelta_{L0}(-\vk')} \rangle . \;\;\;
\label{consistency-delta-2}
\eeqa
This relation can serve as a basis to derive consistency relations for the squeezed limit
of the $n+1$ density correlations (i.e. the limit $k'\rightarrow 0$) 
if we obtain an explicit expression for the response function in the right-hand side.
It turns out that this is possible because the response of the matter distribution to a long-wavelength
mode $\tdelta_{L0}(\vk')$ takes a simple form in the limit $k'\rightarrow 0$
\cite{Kehagias2013,Peloso2013,Creminelli2013}.
Such a change $\Delta \delta_{L0}$ of the initial condition is associated with a change of both 
the linear density and velocity fields, because we change the linear growing mode
where the density and velocity fields are coupled \cite{Kehagias2013},
\beqa
&& \delta_L(\vq,\tau) \rightarrow \hat\delta_L = \delta_L + D_+(\tau) \Delta\delta_{L0} , 
\nonumber \\
&& \vv_L(\vq,\tau) \rightarrow \hat{\vv}_L = \vv_L - \frac{dD_+}{d\tau} 
\nabla_{\vq}^{-1} \cdot \Delta\delta_{L0} . 
\label{change-initial}
\eeqa
Then, in the limit $k'\rightarrow 0$ for the support of $\Delta \delta_{L0}(\vk')$,
the trajectories of the particles are simply modified as \cite{Valageas2014a}
\beq
\vx(\vq,\tau) \rightarrow \hat\vx(\vq,\tau) = \vx(\vq,\tau) + D_+(\tau) \Delta \Psi_{L0}(\vq) ,
\label{translation-1}
\eeq
where $\vq$ is the Lagrangian coordinate of the particles and $\Psi_{L0}$ is the linear 
displacement field,
\beq
\Delta \Psi_{L0} = - \nabla_{\vq}^{-1} \cdot \Delta \delta_{L0}  , \;\;\;
\vx_L(\vq,\tau) = \vq + \Psi_L .
\eeq
The transformation (\ref{translation-1}) simply means that in the limit $k' \rightarrow 0$
smaller-scale structures are displaced by the uniform translation $\Psi_{L0}$
as all particles fall at the same rate in the additional constant force field
$\Delta \vF \propto \nabla_{\vq}^{-1} \cdot \Delta\delta_{L0}$.
In other words, in the limit $k' \rightarrow 0$ we add an almost constant force perturbation
(i.e., a change of the gravitational potential that is linear over $\vq$ for small-scale subsystems)
that gives rise to a uniform displacement, thanks to the weak equivalence principle
\cite{Creminelli2013,Valageas2014a}.
Then, the density field $\delta(\vx,\tau)$ at time $\tau$ is merely displaced by the shift
$D_+(\tau) \Delta \Psi_{L0}$, which gives in Fourier space
\beqa
\tdelta(\vk,\tau) \rightarrow \hat{\tdelta}(\vk,\tau)  & = & \tdelta(\vk,\tau) \, 
e^{-\ii \vk \cdot D_+\Delta\Psi_{L0}} \nonumber \\
&& \hspace{-1cm} = \tdelta(\vk,\tau) - \ii D_+ (\vk\cdot\Delta\Psi_{L0}) \tdelta(\vk,\tau) , \;\;\;
\label{shift-delta-k}
\eeqa
where in the last expression we expanded up to linear order over $\Delta\Psi_{L0}$.
The reader may note that in Eq.(\ref{shift-delta-k}) 
we do not see the additive effect seen at the linear level in the first 
Eq.(\ref{change-initial}). This is because, although the small change of the mean overdensity
over a small structure also leads to a faster (or slower) collapse and distorts the small-scale
clustering, this is a higher-order effect than the kinematic effect studied in this Letter
\cite{Valageas2014b,Kehagias2014}.
Indeed, we shall check in Eqs.(\ref{Ddelta-DdeltaL0}) and (\ref{consistency_relation_deltas})
that this kinematic effect gives rise to factors $\sim 1/k'$ that diverge as $k'\rightarrow 0$.
This is because the linear displacement field is proportional to the inverse gradient of the linear
density field, $\Psi_L = - \nabla_{\vq}^{-1} \delta_L$.
In contrast, the distortions of the small-scale structure (i.e., changes to the shape and amplitude
of the small-scale clustering) are higher-order effects and do not exhibit this factor $1/k'$
\cite{Valageas2014b,Kehagias2014}.
Using the expression
$\Psi_{L0}(\vq) = \int d\vk \, e^{\ii \vk\cdot\vq} \, \ii \frac{\vk}{k^2} \tdelta_{L0}(\vk)$,
we obtain
\beq
k' \rightarrow 0 : \;\;\; \frac{\cD \Psi_{L0}(\vq)}{\cD \tdelta_{L0}(\vk')} 
= \ii \frac{\vk'}{k'^2} , \;\;\; \frac{\cD \tdelta(\vk)}{\cD \tdelta_{L0}(\vk')} 
= D_+ \frac{\vk\cdot\vk'}{k'^2} \tdelta(\vk) .
\label{Ddelta-DdeltaL0}
\eeq
Using this result in the relation (\ref{consistency-delta-2}) gives
\beqa
&& \hspace{-1cm} \langle \tdelta(\vk',\tau') \tdelta(\vk_1,\tau_1) \dots \tdelta(\vk_n,\tau_n) 
\rangle_{k'\rightarrow 0}'  = - P_L(k',\tau') \nonumber \\
&& \times \langle \tdelta(\vk_1,\tau_1) \dots \tdelta(\vk_n,\tau_n) \rangle' 
\sum_{i=1}^n \frac{D_+(\tau_i)}{D_+(\tau')} \frac{\vk_i \cdot \vk'}{k'^2} , \;\; 
\label{consistency_relation_deltas}
\eeqa
which is the density consistency relation in the subhorizon Newtonian regime
\cite{Kehagias2013,Peloso2013,Creminelli2013,Kehagias2014c,Peloso2014,Creminelli2014a,Valageas2014a,Horn2014,Horn2015}.
Here the prime in $\langle \dots \rangle'$ denotes that we removed the Dirac factors
$\delta_D(\sum \vk_i)$.
The remarkable property of Eq.(\ref{consistency_relation_deltas}) is that it does not require 
the wave numbers $\vk_i$ to be in the linear or perturbative regimes. In particular, it still applies 
when $\vk_i$ are in the highly nonlinear regime governed by shell-crossing effects and affected 
by baryonic and galactic processes such as star formation and cooling. 
In fact, under the approximation of the ``squeezed limit'', the long wavelength fluctuation 
$\tilde{\delta}_{L0}(\vk')$ merely transports the small-scale structure of the system. 
This also leads to an another key property of Eq.(\ref{consistency_relation_deltas}), 
namely that it vanishes at equal times, $\tau_1 = .. = \tau_n$. 

\textit{Consistency relations  for velocity and momentum fields:}
\label{sec:consistency_rel-v-p}
%
%
The leading-order effect of a long wavelength 
perturbation is to move smaller structures by an uniform shift and
single-time statistics that only probe the density field cannot see any effect. However,
it is clear that we may detect an effect if we consider the velocity field, as the latter is 
again displaced but also has its amplitude modified.
Thus, the transformation law (\ref{shift-delta-k}) becomes
\beqa
&& \tilde{\vv}(\vk,\tau) \rightarrow \hat{\tilde\vv}(\vk,\tau) = \tilde\vv(\vk,\tau) - \ii D_+ 
(\vk \cdot \Delta\Psi_{L0}) \tvv(\vk,\tau) \nonumber \\
&& \hspace{3.5cm} + \frac{d D_+}{d\tau} \Delta\Psi_{L0} \, \delta_D(\vk) ,
\label{trans-v-k}
\eeqa
where the last factor is the new term, as compared with 
Eq.(\ref{shift-delta-k}), that is associated with the shift of the amplitude.
This yields
\beq
k' \rightarrow 0 : \;\;\; \frac{\cD \tilde\vv(\vk)}{\cD \tdelta_{L0}(\vk')} 
= D_+ \frac{\vk\cdot\vk'}{k'^2} \tilde\vv(\vk) +  \frac{d D_+}{d\tau} \ii \frac{\vk'}{k'^2} \delta_D(\vk) .
\label{Dv-DdeltaL0}
\eeq
Using again the general relation (\ref{C1n-R1n-Fourier}), as in Eq.(\ref{consistency-delta-2})
but where the quantities $\{\rho_1,..,\rho_n\}$ are a combination of density contrasts and
velocities, we obtain 
\beqa
&& \hspace{-0.4cm} \langle \tdelta(\vk',\tau') \prod_{j=1}^n \tdelta(\vk_j,\tau_j) \!
\prod_{j=n+1}^{n+m} \!\!\! \tvv(\vk_j,\tau_j) \rangle_{k'\rightarrow 0}'  = - P_L(k',\tau') \nonumber \\
&& \times \biggl \lbrace \langle \prod_{j=1}^n \tdelta(\vk_j,\tau_j) \!\! \prod_{j=n+1}^{n+m} 
\!\! \tvv(\vk_j,\tau_j) \rangle' 
\sum_{i=1}^{n+m} \frac{D_+(\tau_i)}{D_+(\tau')} \frac{\vk_i \cdot \vk'}{k'^2} \nonumber \\
&& \hspace{0.5cm} + \sum_{i=n+1}^{n+m} \langle \prod_{j=1}^n \tdelta(\vk_j,\tau_j) 
\prod_{j=n+1}^{i-1} \! \tvv(\vk_j,\tau_j) \times \nonumber \\
&& \hspace{0.5cm} \left(  \frac{(d D_+/d\tau)(\tau_i)}{D_+(\tau')} \ii \frac{\vk'}{k'^2} 
\delta_D(\vk_i) \! \right) \prod_{j=i+1}^{n+m} \!\! \tvv(\vk_j,\tau_j)  \rangle' \biggl \rbrace .
\label{consistency_relation_v}
\eeqa
If we take $\vk_i \neq 0$, as usual for studies of Fourier-space polyspectra,
the last term vanishes and we recover the same form as for the consistency relation 
(\ref{consistency_relation_deltas}) of the density field.
However, this new Dirac term will give a nonzero contribution in configuration space.
Therefore, real-space correlation functions will obey consistency relations that differ
from those of the density field if we include cross-correlations with the velocity field.
The correlation functions in Eq.(\ref{consistency_relation_v}) are $3^m$-component quantities,
as the velocity field is a 3-component vector. One may obtain scalar relations by taking
for instance the divergence of the velocity field or considering the 
components along cartesian coordinates.
The divergence $\theta=\nabla\cdot\vv$ was considered in \cite{Peloso2013,Peloso2014}. 
We recover the fact that it obeys relations similar to the density field because the new Dirac term
$\delta_D(\vk_i)$ disappears as $\tilde\theta_i = \ii \vk_i\cdot\tvv_i$.
We will rather focus on the divergence of the momentum field in this Letter,
as it yields new terms in the consistency relations and it satisfies a direct relationship 
with the density field which may provide useful checks.


One simple way to make the last term in Eq.(\ref{trans-v-k}) relevant in Fourier space
at nonzero wave numbers is to consider composite operators,
that is, products of the velocity field with other fields. 
Therefore, we define the momentum $\vp$ as
\beq
\vp = (1+\delta) \vv ,
\label{p-def}
\eeq
which reads in Fourier space as
\beq
\tvp(\vk) = \tvv(\vk) + \int d\vk_1 d\vk_2 \; \delta_D(\vk_1+\vk_2-\vk) \tdelta(\vk_1) \tvv(\vk_2) .
\label{p-Fourier-def}
\eeq
Using Eqs.(\ref{Ddelta-DdeltaL0}) and (\ref{Dv-DdeltaL0}) we obtain
\beqa
k' \rightarrow 0 : \;\;\; \frac{\cD \tilde\vp(\vk)}{\cD \tdelta_{L0}(\vk')} 
& = & D_+ \frac{\vk\cdot\vk'}{k'^2} \tilde\vp(\vk) \nonumber \\
&& \hspace{-1cm} +  \frac{d D_+}{d\tau} \ii \frac{\vk'}{k'^2} [ \delta_D(\vk) + \tdelta(\vk) ] . \hspace{1cm}
\label{Dp-DdeltaL0}
\eeqa
The first term, which is common with Eqs.(\ref{Ddelta-DdeltaL0}) and (\ref{Dv-DdeltaL0}), 
corresponds to the translation of the system, whereas the second term corresponds to the 
additional velocity generated by the long-wavelength mode.
Thanks to the convolution in Eq.(\ref{p-Fourier-def}) it is now nonzero for $\vk \neq 0$.
However, in contrast to the translation term, it transforms the field because the functional derivative
of the momentum $\tilde\vp$ now gives rise to a factor that is proportional to the density contrast
$\tdelta$.
In a fashion similar to Eqs.(\ref{consistency_relation_deltas}) and (\ref{consistency_relation_v}), 
we obtain the consistency relation
\beqa
&& \hspace{-0.3cm} \langle \tdelta(\vk',\tau') \prod_{j=1}^n \tdelta(\vk_j,\tau_j) \!\!
\prod_{j=n+1}^{n+m} \!\! \tvp(\vk_j,\tau_j) \rangle_{k'\rightarrow 0}'  = - P_L(k',\tau') \nonumber \\
&& \times \biggl \lbrace \langle \prod_{j=1}^n \tdelta(\vk_j,\tau_j) \!\! \prod_{j=n+1}^{n+m} 
\!\! \tvp(\vk_j,\tau_j) \rangle' 
\sum_{i=1}^{n+m} \frac{D_+(\tau_i)}{D_+(\tau')} \frac{\vk_i \cdot \vk'}{k'^2} \nonumber \\
&& \hspace{0.5cm} + \sum_{i=n+1}^{n+m}  \frac{(d D_+/d\tau)(\tau_i)}{D_+(\tau')}
\langle \prod_{j=1}^n \tdelta(\vk_j,\tau_j) 
\prod_{j=n+1}^{i-1} \tvp(\vk_j,\tau_j) \nonumber \\
&& \hspace{0.5cm} \times \left( \ii \frac{\vk'}{k'^2} [ \delta_D(\vk_i) + \tdelta(\vk_i,\tau_i) ] \! \right) \! 
\prod_{j=i+1}^{n+m} \! \tvp(\vk_j,\tau_j)  \rangle' \biggl \rbrace .
\label{consistency_relation_p}
\eeqa
The first term, which has the same form as the density and velocity consistency relations
(\ref{consistency_relation_deltas}) and (\ref{consistency_relation_v}), is due to the translation
of smaller-scale structures by the long wavelength mode $k'$. The new second term
is due to the additional velocity and arises from the second term in Eq.(\ref{Dp-DdeltaL0}).
This term has a different form as it transforms one small-scale momentum mode,
$\tvp(\vk_i)$, into a small-scale density mode, $\tdelta(\vk_i)$.
Moreover, this new term no longer automatically vanishes at equal times.
This leads to a nontrivial consistency relation at equal times,
when $\tau'=\tau_1=..=\tau_{n+m}$,
\beqa
&& 
\langle \tdelta(\vk') \prod_{j=1}^n \tdelta(\vk_j) \!
\prod_{j=n+1}^{n+m} \tvp(\vk_j) \rangle_{k'\rightarrow 0}'  = - \ii \, P_L(k') \frac{d\ln D_+}{d\tau}
\nonumber \\
&& \times \sum_{i=n+1}^{n+m} \langle \prod_{j=1}^n \! \tdelta(\vk_j) \!
\prod_{j=n+1}^{i-1} \! \tvp(\vk_j) \left( \frac{\vk'}{k'^2} [ \delta_D(\vk_i) + \tdelta(\vk_i) ] \right)  \!
\nonumber \\
&& \times \prod_{j=i+1}^{n+m} \!\! \tvp(\vk_j)  \rangle' 
\label{consistency_relation_p-single-time}
\eeqa
where we did not write the common time $\tau$ of all fields.
We can also obtain a consistency relation that involves both the density and velocity
fields $\tdelta$ and $\tvv$, together with the momentum field $\tvp$, and it shows the same
behaviors.


To obtain a scalar quantity from the momentum field $\vp$ we consider its divergence,
\beq
\lambda \equiv \nabla \cdot \left[ (1+ \delta) \vv \right] , \;\;\;
\tlambda(\vk) = \ii \vk \cdot \tvp(\vk) .
\eeq
Then, the consistency relation for the divergence $\tlambda$ follows from 
Eq.(\ref{consistency_relation_p}). This gives
\beqa
&& \hspace{-0.5cm} \langle \tdelta(\vk',\tau') \prod_{j=1}^n \tdelta(\vk_j,\tau_j) \!\!
\prod_{j=n+1}^{n+m} \!\! \tlambda(\vk_j,\tau_j) \rangle_{k'\rightarrow 0}'  = - P_L(k',\tau') \nonumber \\
&& \times \biggl \lbrace \langle \prod_{j=1}^n \tdelta(\vk_j,\tau_j) \! \prod_{j=n+1}^{n+m} \!
\tlambda(\vk_j,\tau_j) \rangle' 
\sum_{i=1}^{n+m} \frac{D_+(\tau_i)}{D_+(\tau')} \frac{\vk_i \cdot \vk'}{k'^2} \nonumber \\
&& \hspace{0.5cm} - \sum_{i=n+1}^{n+m} \langle \tdelta(\vk_i,\tau_i) 
\prod_{j=1}^n \tdelta(\vk_j,\tau_j) \prod_{\substack{j=n+1 \\ j\neq i}}^{n+m} \tlambda(\vk_j,\tau_j) 
\rangle' \nonumber \\
&& \hspace{1cm} \times \frac{(d D_+/d\tau)(\tau_i)}{D_+(\tau')} \frac{\vk_i\cdot\vk'}{k'^2} 
\biggl \rbrace .
\label{consistency_relation_lambda}
\eeqa
At equal times this gives the relation
\beqa
&& \hspace{-1cm} \langle \tdelta(\vk') \prod_{j=1}^n \tdelta(\vk_j) 
\prod_{j=n+1}^{n+m} \tlambda(\vk_j) \rangle_{k'\rightarrow 0}'  = P_L(k') \frac{d \ln D_+}{d\tau}
\nonumber \\
&& \hspace{-0.2cm} \times \sum_{i=n+1}^{n+m} \frac{\vk_i\cdot\vk'}{k'^2} 
\langle \tdelta(\vk_i) 
\prod_{j=1}^n \tdelta(\vk_j) \prod_{\substack{j=n+1 \\ j\neq i}}^{n+m} \tlambda(\vk_j) 
\rangle' , \;\;\;
\label{consistency_relation_lambda-single-time}
\eeqa
where we did not write the common time $\tau$ of all fields.
We can easily check the relation (\ref{consistency_relation_lambda}) by noticing that the divergence
$\lambda$ is related to the density field through the continuity equation,
$\frac{\partial\delta}{\partial\tau} + \nabla \cdot \left[ (1+ \delta) \vv \right] = 0,$ 
which implies 
$\lambda = - \partial\delta / \partial\tau$.
Therefore, Eq.(\ref{consistency_relation_lambda}) can be directly obtained from the density
consistency relation (\ref{consistency_relation_deltas}) by taking partial derivatives with respect
to the times $\tau_j$.

\textit{Applications:}
\label{sec:applications}
%
%
As for the density contrast relation (\ref{consistency_relation_deltas}), the new consistency
relations that we have obtained in this Letter
are valid beyond the perturbative regime, after shell crossing, and also apply to baryons,
gas and galaxies, independently of the bias of the objects that are used.
Indeed, they only use the property (\ref{translation-1}), which states that at leading order  
the effect of a long-wavelength mode is to move smaller scale structures without disturbing
them. This relies on the equivalence principle, which states that all particles (and astrophysical
objects) fall at the same rate in a gravitational potential well (the inertial mass is also the
gravitational mass) \cite{Creminelli2013,Valageas2014a,Horn2014}.

After shell crossing we enter a multi-streaming regime where the velocity field is multi-valued:
at a given position there are several streams with different velocities as they cross
each other and build a nonzero velocity dispersion, as within virialized halos.
Nevertheless, our results remain valid. In that case, $\vv$ can be taken as any of these streams
or as any given linear combination of them, because all stream velocities are modified in the same
way.
In multi-streaming regions, such as high-density non-linear environments like clusters or
filaments, it is more practical to work with the mean momentum $\vp$, where
Eq.(\ref{p-def}) reads in the case of several streams $i$ as
$\vp = \sum_{\rm streams} (1+\delta_i) \vv_i$,
or in terms of a phase-space distribution function as
$\vp = \int d\vv f(\vx,\vv) \vv$ .
This is also more convenient for observational purposes as we only observe velocities where
there is baryonic matter, so that it is easier to build momentum maps than velocity maps,
which are difficult to measure in voids.
The expression (\ref{Dp-DdeltaL0}) remains valid in these multi-streaming regions, as the
first term simply expresses the translation of the smaller-scale system while the second term
expresses the large-scale constant additive term that is added to all velocities. 
Thus, these consistency relations only rely on
\begin{enumerate}
\item Gaussian initial conditions (Eq.(\ref{C1n-R1n-Fourier}));
\item The weak equivalence principle (Eq.(\ref{translation-1})).
\end{enumerate}
Therefore, a detection of a violation would be a signature of non-Gaussian initial conditions
or of a modification of gravity (or a fifth force).
In practice, we also need to make sure the large-scale wave number $k'$ is within the linear regime
and far below the other wave numbers $k_i$, so that the limit $k'\rightarrow 0$
is reached.


The simplest relation that does not vanish at equal times
is the bispectrum with one momentum field. 
From Eqs.(\ref{consistency_relation_p-single-time}) and 
(\ref{consistency_relation_lambda-single-time}) we obtain for $\vk \neq 0$
\beq
\langle \tdelta(\vk') \tdelta(\vk) \tvp(-\vk) \rangle_{k'\rightarrow 0}'  
= - \ii \frac{\vk'}{k'^2} \frac{d\ln D_+}{d\tau} P_L(k') P(k) \nonumber
\eeq
\beq 
\langle \tdelta(\vk') \tdelta(\vk) \tlambda(-\vk) \rangle_{k'\rightarrow 0}'  
= -\frac{\vk\cdot\vk'}{k'^2} \frac{d\ln D_+}{d\tau} P_L(k') P(k).\nonumber
\eeq
Here $P(k)$ is the non-linear density power spectrum and  these
relations remain valid in the nonperturbative non-linear regime.
For galaxies these relations are
\beq
\langle \tdelta(\vk') \tdelta_g(\vk) \tvp_g(-\vk) \rangle_{k'\rightarrow 0}'  
= - \ii \frac{\vk'}{k'^2} \frac{d\ln D_+}{d\tau} P_L(k') P_{\delta_g \delta_g}(k) 
\label{bispectrum_p-galaxies}
\eeq
\beq 
\langle \tdelta(\vk') \tdelta_g(\vk) \tlambda_g(-\vk) \rangle_{k'\rightarrow 0}'  
= -\frac{\vk\cdot\vk'}{k'^2} \frac{d\ln D_+}{d\tau} P_L(k') P_{\delta_g \delta_g}(k)
\label{bispectrum_lambda-galaxies}
\eeq
where $\tdelta$ and $P_L$ are again the matter density field and linear power spectrum, 
$\tdelta_g$ and $\tvp_g$ the galaxy density contrast and momentum, and $P_{\delta_g \delta_g}$
the galaxy density power spectrum.
In Eqs.(\ref{bispectrum_p-galaxies})-(\ref{bispectrum_lambda-galaxies}) we kept the long mode
$k'$ as the matter density contrast $\tdelta$ because the actual consistency relation is with
respect to the initial condition $\delta_{L0}$, as in Eq.(\ref{consistency-delta-1}), 
and $\delta(\vk')$ merely stands for $D_+(\tau') \delta_{L0}(\vk')$ in the limit $k'\rightarrow 0$.
If we wish to write Eqs.(\ref{bispectrum_p-galaxies})-(\ref{bispectrum_lambda-galaxies})
in terms of galaxy fields only, we need to assume that the matter and galaxy density fields
are related by a finite bias $b_1$ in the limit $k' \rightarrow 0$. Then, 
Eq.(\ref{bispectrum_lambda-galaxies})  becomes
\beqa 
b_1 \langle \tdelta_g(\vk') \tdelta_g(\vk) \tlambda_g(-\vk) \rangle_{k'\rightarrow 0}'  
& = & -\frac{\vk\cdot\vk'}{k'^2} \frac{d\ln D_+}{d\tau} P_{\delta_g\delta_g}(k') \nonumber \\
&& \times P_{\delta_g \delta_g}(k) , 
\label{bispectrum_lambda-galaxies-bias}
\eeqa
where we assumed a deterministic large-scale limit $b_1$ for the galaxy bias,
$k' \rightarrow 0 : \delta_g(\vk') = b_1 \, \delta(\vk')$ .
Then, Eq.(\ref{bispectrum_lambda-galaxies-bias}) can be used as a measurement of the
large-scale bias $b_1$.

\textit{Conclusions:}
\label{sec:conclusion}
We have obtained in this Letter very general and exact consistency relations
for cosmological structures that do not vanish at equal times by taking cross correlations
with the velocity or momentum fields. These relations, which are nonperturbative and also
apply to galaxy fields, could be useful to constrain the Gaussianity of the initial conditions, 
deviations from General Relativity, or the large-scale galaxy bias.
%
\begin{acknowledgments}
This work is supported in part by the French Agence Nationale de la Recherche
under Grant ANR-12-BS05-0002. DFM is supported by the Norwegian Research Council.
\end{acknowledgments}
\bibliography{ref1}   
\end{document}